\documentclass{article}
\usepackage{epsfig}

\begin{document}
\title{\bf The role of curvature in quantum statistical mechanics}

\author{Marcelo R. Ubriaco\thanks{Electronic address:ubriaco@ltp.uprrp.edu}}
\date{Laboratory of Theoretical Physics\\Department of Physics\\University of Puerto Rico\\17 Ave. Universidad Ste. 1701\\
San Juan, PR 00925-2537, USA}

\maketitle

\begin{abstract}
In this manuscript, we calculate the scalar curvature of a two-dimensional thermodynamic space to study
the properties of two thermodynamic systems. In particular, we study the stability and possible anyonic behavior 
of quantum group invariant systems and systems
with fractal distribution functions.

\end{abstract}

\section{Introduction}
During the last five decades there has been a lot of development in using geometry to study some properties of thermodynamic systems \cite{Ti}-\cite{AN}.
In particular, it consists in defining a metric in a two dimensional parameter space and calculate the corresponding scalar curvature
as a measure of the correlations strength of the system \cite{IJKK}-\cite{R}, with applications to classical and quantum gases
\cite{R1}\cite{NS}\cite{JM1}\cite{BH}, magnetic systems \cite{JM2}-\cite{JJK}, non-extensive statistical mechanics \cite{T-B}-\cite{O},
anyon gas, fractional statistics and deformed boson and fermion systems \cite{MH1}, systems
with fractal distribution functions \cite{MRU1}, quantum group invariant systems \cite{MRU2}, systems with $M$-statistics \cite{MRU3}, and  those related to Dunkl differential-difference operators \cite{MRU4}.  Some of the basic 
 results of these approaches include the relationships between the metric  
 with the correlations of the  stochastic variables, and  the scalar curvature $R$ 
 with the stability of the system, and the facts that  the scalar curvature $R$ vanishes for the
 classical ideal gas,   $R>0 (R<0)$ for a boson (fermion) ideal gas, and it is singular at a critical point.
In Section \ref{Geom} we give a general but brief introduction to the calculation of the thermodynamic metric and the corresponding scalar curvature. In Sections
\ref{QG} and \ref{Fractal} we apply the formalism to study the stability and possible anyonic behavior for the cases of quantum group invariant systems  and systems with fractal distribution functions, respectively. In constrast to the quantum group case  the fractal density distribution is not exponential.
\section {Thermodynamic curvature}\label{Geom}
In order to make the discussion as general as possible, we start with a general probability  density $\rho$ which is not necessarily exponential. Therefore, let us define
\begin{equation}
\rho=\frac{1}{Z}\prod_{l=0}F_l.
\end{equation}
Starting with the relative entropy for two close densities
$\rho(\beta^{\alpha})$ and $\rho(\beta_{\alpha}+ d\beta_{\alpha})$, where
$\beta_1=\beta$ and $\beta_2=-\beta\mu$, (two-dimensional coordinate space). The metric is defined \cite{IJKK} as the second order term
in the expansions of the information distance
\begin{equation}
I=Tr\rho\left(\ln \rho(\beta^{\alpha})-\ln \rho(\beta^{\alpha}+d\beta^{\alpha})\right).
\end{equation}
Expanding, at second order we obtain the metric

\begin{equation}
g_{\alpha\gamma}=\frac{\partial^2\ln Z}{\partial  \beta_{\alpha}\partial\beta_{\gamma}}-Tr\left(\rho\sum_{l=0}\frac{\partial^2\ln F_l}{\partial  \beta_{\alpha}\partial\beta_{\gamma}}\right),\label{metric}
\end{equation}
where the second term vanishes for exponential distributions with an exponent linear in the inverse temperature $\beta$.
The scalar curvature $R$ is given by
\begin{equation}
R=\frac{2}{det g}R_{1212},
\end{equation}
where the Riemann tensor reduces to 
\begin{equation}
R_{\alpha\beta\gamma\delta}=g^{\eta\theta}(\Gamma_{\eta\gamma\lambda}\Gamma_{\theta\beta\gamma}-\Gamma_{\eta\alpha\gamma}\Gamma_{\theta\beta\lambda})
\end{equation}
The calculation of $R$ simply reduces to solve the following determinant
\begin{equation}
 R=\frac{1}{2 (detg)^2}\left|\begin{array}{ccc} g_{11} & g_{22} & g_{12}  \\ g_{11,1} & g_{22,1} & g_{21,1} \\
g_{11,2} & g_{22,2}& g_{21,2} \end{array}\right|,
\end{equation}
where $g_{\alpha\beta,\lambda}=\partial g_{\alpha\beta}/\partial\beta_{\lambda}$. The scalar curvature values
give us the following information:
 Classical case: $ R=0$, Bosons: $ R>0$, Fermions :$ R<0$, at a phase transition: $R\rightarrow \infty$, and $R\approx 0$ implies a more stable system.

\section {Geometry of Quantum Group invariant systems} \label{QG}
In the last twenty years there has been considerable interest on applications of quantum groups
\cite{Jimbo}\cite{Ch}, in addition to the theory of integrable models, to diverse areas of theoretical physics.
The vast published literature on this subject includes formulations of quantum group versions of Lorentz and Poincar\'{e}
algebras \cite{CW}, its use as internal quantum symmetries in quantum mechanics and field theories \cite{AV}, molecular
and nuclear physics \cite{I}, and  the formulation
and study on the implications of imposing  quantum group invariance in thermodynamic systems \cite{MRU5}. In particular, it has been shown \cite{MRU6} that quantum group gases exhibit anyonic behavior in two and three dimensions. 
 More recent
applications include gravity theories wherein the discreteness and non-commutative properties 
of space-time  at the Planck scale are approached mathematically  by replacing the local symmetry by a quantum group
symmetry \cite{MN} and applications to the phenomena of  entanglement \cite{KWL}. 
In particular,we focuses on the quantum group $SU_q(2)$ which consists of the set of matrices $T=\left(\begin{array}{cc} a & b \\ c & d\end{array}\right)$
with elements $\{a,b,c,d\}$ generating the algebra

$$ab=q^{-1}ba  ,\;\; ac=q^{-1}ca,\;\;bc=cb  , \;\; dc=qcd,\;\;db=qbd  , $$
$$  da-ad=(q-q^{-1})bc,\;\; det_{q}T\equiv ad-q^{-1}bc=1 ,$$

with the unitary conditions \cite{VWZ} $\overline{a}=d, \overline{b}=q^{-1}c$
and $q\in {\bf R}$. Hereafter, we take $0< q<\infty$.
\subsection{$SU_q(2)$-bosons}
The $SU_q(2)$-invariant bosonic algebra is given by the following relations:

$$\Phi_2\overline{\Phi}_2-q^2\overline{\Phi}_2\Phi_2=1,\;\;
\Phi_1\overline{\Phi}_1-q^2\overline{\Phi}_1\Phi_1=1+(q^2-1)\overline{\Phi}_2\Phi_2,$$
$$\Phi_2\Phi_1=q\Phi_1\Phi_2,\;\;
\Phi_2\overline{\Phi}_1=q\overline{\Phi}_1\Phi_2 .$$

The simplest Hamiltonian written in terms of the operators $\Phi_j$ is simply written as
\begin{equation}
{\cal H}_B=\sum_\kappa \varepsilon_\kappa({\cal N}_{1,\kappa}+{\cal N}_{2,\kappa}),\label{H}
\end{equation}
where $[\overline{\Phi}_{i,\kappa},\Phi_{\kappa',j}]=0$ for
$\kappa\neq\kappa'$ and ${\cal N}_{i,\kappa}=\overline{\Phi}_{i,\kappa}\Phi_{i,\kappa}$. 
For a given $\kappa$ the $SU_q(2)$ bosons are written 
in terms of boson operators $\phi_{i}$
and $\phi_{i}^{\dagger}$ with usual commutation relations:
$[\phi_i,\phi_j^{\dagger}]=\delta_{ij}$ as follows
$$\Phi_j=(\phi_j^\dagger)^{-1} \{N_j\}q^{N_{j+1}},\;\;\;
\overline{\Phi}_j=\phi_j^\dagger q^{N_{j+1}}, \;\;\; j=1,2$$

leading to the  interacting boson Hamiltonian
\begin{equation}
{\cal H}_B=\sum_{\kappa}\frac{\varepsilon_{\kappa}}{q^2-1}\sum_{m=1}^{\infty}
\frac{2^m \ln^mq}{m!}\left(N_{1,\kappa}+N_{2,\kappa}\right)^m,
\end{equation}
where $N_{i,\kappa}$ is the ordinary boson number operator and the $q$-number\\
 $\{n\}=\frac{1-q^{2n}}{1-q}$. Equation (\ref{H}) becomes the standard free boson hamiltonian at $q=1$.

 The grand partition function ${\cal Z}_B$ is given by
\begin{eqnarray}
{\cal Z}_B&=&Tr \;e^{-\beta\sum_{\kappa}\varepsilon_{\kappa}({\cal N}_{1,\kappa}+{\cal N}_{2,\kappa})}\nonumber\\
&=&\prod_\kappa\sum_{n=0}^{\infty}\sum_{m=0}^{\infty}
e^{-\beta\varepsilon_\kappa\{n+m\}}e^{\beta\mu(n+m)},\label{Zb}
\end{eqnarray}
which after rearrangement simplifies to the Equation
\begin{equation}
{\cal Z}_B=\prod_\kappa\sum_{m=0}^{\infty} (m+1)e^{-\beta\varepsilon_{\kappa}\{m\}}z^m,\label{Z}
\end{equation}
where $z=e^{\beta\mu}$ is the fugacity. 
Since the density distribution is exponential, the second term in Equation (\ref{metric}) vanishes. The scalar curvature for $D=2,3$ is given by the Equation
\begin{equation}
R_D=\frac{(D+4\nu) \pi^{\nu}\lambda^D}{V_D}\left(\frac{c_{\nu}b^2_{\nu}+a_{\nu}d_{\nu}b_{\nu}-2a_{\nu}c^2_{\nu}}{[(D+4\nu)a_{\nu}c_{\nu}-(D+2\nu-1)b^2_{\nu}]^2}\right),\label{RQGB}
\end{equation}
where $ \nu=\frac{D-2}{2}$ and for example
\begin{equation}
a_{\nu}=\int_0^{\infty}x^{\nu}\ln \left(1+\sum_{n=1}(n+1)e^{-x\{n\}}z^n\right) dx,
\end{equation}

where $b_{\nu},\;c_{\nu}$ and $d_{\mu}$ are obtained by taking the corresponding derivatives with respect to $\beta_1$ and $\beta_2$.
Figures 1  displays the results of a numerical calculation of Equation (\ref{RQGB}) for the scalar curvature of a QGB system in $D=3$.
Figure 1 shows that for $0<q\leq 1$ the curvature is positive for all $z$ values, and therefore the system is bosonic. In addition, the system becomes  more stable than the standard Bose-Einstein case at higher values of $z$. For $1<q\approx 1.2$ the system exhibits anyonic behavior as a function of $z$ and becomes  purely fermionic for all values of $z$ and $q>1.2$. 
 The calculation of the Equation (\ref{RQGB}) for $D=2$ shows that for $0<q<1$ the system is bosonic and more stable at $z\approx 1$ than the standard case $q=1$. For $q>1$ the system goes from bosonic to  fermionic and becomes bosonic becoming more unstable as $z$ approaches the value of $1$.

\begin{figure}
\begin{center}
\epsfig{file= 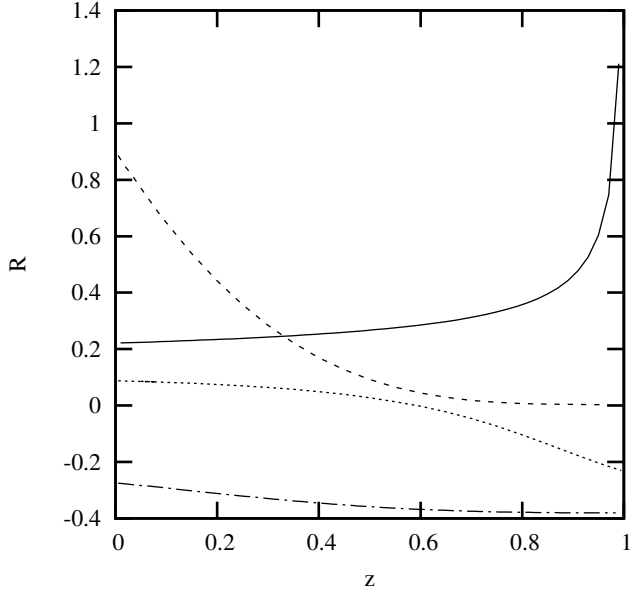,bbllx=50pt,bblly=120pt,bburx=430pt,bbury=250pt}
\end{center}
\caption[]{ The scalar curvature $R$, in units of $\lambda^3V^{-1}$, as a function of the fugacity $z$ for quantum group bosons at $D=3$ and constant $\beta$ for the cases of $q=1$ (solid
line), $q=0.5$ (dashed line) and $q=1.15$ (dotted line) and $q=2$ (dashed-dotted line).}
\end{figure}
\newpage
\subsection{$SU_q(2)$-fermions}
The $SU_q(2)$ invariant fermionic algebra is defined by the set of relations

$$\{\Psi_{2},\overline{\Psi}_{2}\}=1,\;\;\;
\{\Psi_{1},\overline{\Psi}_{1}\}=1 - (1-q^{-2})\overline{\Psi}_{2}\Psi_{2}\label{f1}$$ 
$$\Psi_{1}\Psi_{2}=-q \Psi_{2}\Psi_{1},\;\;\; 
\overline{\Psi}_{1}\Psi_{2}=-q \Psi_{2}\overline{\Psi}_{1}$$
$$\{\Psi_{1},\Psi_{1}\}=0=\{\Psi_{2},\Psi_{2}\}.$$
The corresponding Hamiltonian in terms of
the operators $\Psi_i$ is simply 
\begin{equation}
{\cal H}_F=\sum_{\kappa}^{}\varepsilon_{\kappa}({\cal M}_{1,\kappa}+
{\cal M}_{2,\kappa}),\label{h}
\end{equation}
where ${\cal M}_{i\kappa}=\overline{\Psi}_{i,\kappa}\Psi_{i,\kappa}$
and $\{\overline{\Psi}_{\kappa,i},\Psi_{\kappa',j}\}=0$ for $\kappa\neq\kappa'$.
The occupation numbers are restricted to $m=0,1$ and 
therefore $SU_q(N)$ fermions satisfy the Pauli exclusion principle. 
A simple check shows that the $SU_q(2)$ fermionic algebra
is consistent with the following representation of $\Psi_i$ operators
in terms of fermion operators $\psi_j$ 

$$\Psi_2=\psi_2,\;\;\;
\overline{\Psi}_2=\psi_2^{\dagger}$$
$$\Psi_1=\psi_1\left(1+(q^{-1}-1)M_2\right),\;\;
\overline{\Psi}_1=\psi_1^{\dagger}\left(1+(q^{-1}-1)M_2\right)$$
where $M_2=\psi_2^{\dagger}\psi_2$.\\

 From Equation (\ref{h}) we see  that the original Hamiltonian becomes 
 the interacting 
Hamiltonian
\begin{equation}
H_F=\sum_\kappa \varepsilon_\kappa\left(M_{1,\kappa}+M_{2,\kappa}+(q^{-2}-1)
M_{1,\kappa}M_{2,\kappa}\right).\label{HF}
\end{equation}
Therefore the parameter $q\neq 1$ mixes the two
degrees of freedom in a nontrivial way through a quartic term in the Hamiltonian.

The grand partition function ${\cal Z}_F$ is written
\begin{eqnarray}
{\cal Z}_F&=&\prod_\kappa\sum_{n=0}^1\sum_{m=0}^1e^{-\beta\varepsilon_\kappa(n+m
-(1-q^{-2})mn}e^{\beta\mu(n+m)}\nonumber\\
&=&\prod_\kappa \left(1+2e^{-\beta(\varepsilon_\kappa-\mu)}+e^{-\beta\left
(\varepsilon_\kappa(q^{-2} +1)-2\mu\right)}\right) \label{ZF}
\end{eqnarray}
which for $q=1$ becomes the square of a single fermion type grand partition
function. 
Figure 2 is a graph of $R$ vs. $q$ for the values $z=0.1,0.5,2,10$. For $z=0.1$ the system becomes bosonic for $q\geq 2.3$. This QGF system is more stable 
than the $q=1$ system when $q<1$ and it is more unstable than the $q=1$ system when $q>1$.
\begin{figure}
\begin{center}
\epsfig{file= 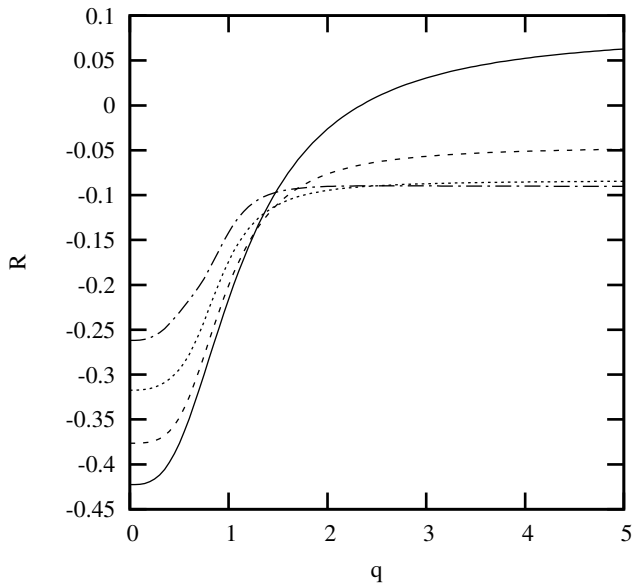,bbllx=50pt,bblly=120pt,bburx=430pt,bbury=250pt}
\end{center}
\caption[]{The scalar curvature $R$  for quantum group fermions at $D=3$, in units of $\lambda^3V^{-1}$, as a function of the parameter $q$ and values for the fugacity $z=0.1$ (solid line),   $z=0.5$ (dashed line),   $z=2$ (dotted line) and $z=10$ (dashed-dotted line).}
\end{figure}
On the other hand, quantum group, $SU_q(2)$, fermions do not exhibit anyonic behavior at $D=2$. However, a calculation of the virial coefficients for the
two parameter case reveals that $SU_{q/p}(2)$ fermions have anyonic behavior \cite{AAA}.
\newpage
\section {Geometry of systems with fractal distribution functions} \label{Fractal}
From the theory of fractals \cite{F} we learned that given a statistical weight $\Omega(q,\delta)$  of a system with
order parameter $q$ and resolution $\delta$, the fractal dimension is defined as the exponent $d=D_q$ which will make the product $lim_{\delta\rightarrow 0}\Omega(q,\delta)\delta^d$ finite.
With use of  the definition of the Boltzmann entropy  $S(q,\delta)=\ln\Omega(q,\delta)$, the relation between the entropy and the fractal dimension $D_q$ is given by
\begin{equation}
D_q=-\lim_{\delta\rightarrow 0}\frac{S(q,\delta)}{\ln\delta}.
\end{equation}
Based on these definitions and with use of the Boltzmann's H theorem, the generalized entropy and distribution functions for
classical and quantum gases were calculated in Ref.\cite{BD}. The average number of particles with energy $\epsilon$ was shown to
be given by
\begin{equation}
<n(\epsilon)>=\frac{1}{\rho_j+a}\label{n},
\end{equation}
where $\rho_j=[1+\beta(q-1)(\epsilon-\mu)]^{1/(q-1)}$ , $a=0$ for the classical case, and the values $a=-1$ and $a=1$ correspond to Bose-Einstein and Fermi-Dirac cases, respectively. For $q=1$, Equation (\ref {n}) becomes the standard textbook result for classical and quantum ideal gases.
The distribution functions in Equation (\ref {n}) were also obtained in Ref.\cite{BDA} by considering a dilute gas approximation to the partition function of a non-extensive statistical mechanics originally proposed in Ref.\cite{T}.
In this Section we calculate the scalar curvature of systems with average particle number according to Equation (\ref {n}). The probability densities that leads to Equation (\ref{n}) were obtain in Ref. \cite{MRU7} and they are given by
\begin{equation}
\rho_{MB}=\frac{1}{Z_{MB}}\prod_{j=0}\frac{1}{n_j!}\rho^{-n_j},
\end{equation}
\newpage
for the classical case, and
\begin{equation}
\hat{\rho}=\frac{1}{Z}\prod_{j=0}\rho^{-\hat{n}_j},
\end{equation}
for the Bose-Einstein and Fermi-Dirac cases. Since the density distribution is not exponential, the second term in Equation (\ref{metric}) does not vanish and the metric for the three cases can be summarized in the general formula \cite{MRU1}
\begin{equation}
g_{\alpha\gamma}=\sum_{l=0}\frac{<n_l>^2}{\rho_l}\frac{\partial\rho_l}{\partial\beta^{\alpha}}\frac{\partial\rho_l}{\partial\beta^{\gamma}}.
\end{equation}

The curvature is given by the expression
\begin{equation}
R=\frac{\lambda^3}{4V (det g)^2}\left(5h_{1/2}h^2_{3/2}-6h^2_{1/2}h_{5/2}+h_{3/2}h_{-1/2}h_{5/2}\right), \label{Rfractal}
\end{equation}

where the function
\begin{equation}
h_{\lambda}=\frac{2}{\sqrt{\pi}(q-1)^{(\lambda+1)}}\frac{\Gamma(\lambda+1)\Gamma(\frac{q}{q-1}-\lambda)}{\Gamma(\frac{2q-1}{q-1})}\frac{1}
{[1+(q-1)\gamma]^{\frac{q}{q-1}-\lambda}}.
\end{equation}
After replacement of the definition of the function $h_{\lambda}$ we get that the scalar curvature for the classical fractal case is identically equal to zero. Therefore, in this case the parameter $q$ does not play any role as far as correlations are concerned. 

For the bosonic and fermionic cases the functions $h_{\lambda}$ are replaced in Equation (\ref{Rfractal}) by the functions
\begin{equation}
G_{\lambda}^{\pm}=\frac{2}{\sqrt\pi}\int_0^{\infty}\frac{x^{\lambda}\Omega^{\frac{3-2q}{q-1}}}{(\Omega^{\frac{1}{q-1}}\pm 1)^2}dx,
\end{equation}
where the $(+)$ sign is for fermions and the $(-)$ sign for bosons, and the function $\Omega=1+(q-1)(x-\beta \mu)$.
 Numerical calculations for the boson and fermion systems show that the corresponding values of $R$ as a function of the fugacity $z$ are closer to zero than those
in the $q=1$ case, implying that the departure from the value $q=1$ makes the systems more stable. Therefore, for $q\neq 1$
bosons will be less attractive and fermions less repulsive that their standard counterparts. Our results also show  that the sign of $R$ remains unchanged as a function of $z$ implying that these systems do not exhibit anyonic behavior, a fact that  looks impossible to check by performing  an expansion for $z\approx 0$ to obtain  the second virial coefficient
because the partition function is a function of $\ln z$.
\\
We can summarize that the two interesting systems are two examples of the usefulness of the calculation of the thermodynamic curvature.  These calculations give us important information about the stability and possible anyonic behavior of the system under study as a function of the temperature and the particular additional parameter involved.
\newpage
\section*{Acknowledgment}
I am very grateful to Professor Abdullah Algin and Osmangazi University Administration for their hospitality and giving me the opportunity of visiting
their beautiful country. I also extend my gratitude to the physics graduate students Hakan Cetinkaya and Erbil Silik.


\begin{thebibliography}{99}
\bibitem{Ti}  Tisza L 1966 {\it Generalized Thermodynamics} (MIT, Cambridge).
\bibitem{GW}  Griffiths R and  Wheeler J 1970 {\it Phys. Rev.} A {\bf 2 }1047.
\bibitem{W}  Weinhold F 1975 {\it J. Chem. Phys.} {\bf 63}  2479.
\bibitem{A1}  Amari S 1985 {\it Differential-Geometrical Methods in Statistics} (Springer-Verlag, Berlin).
\bibitem{AN} Amari S and  Nagaoka H 2000 {\it Methods of Information Geometry} (AMS, Rhode Island).
\bibitem{R1}  Ruppeiner G 1979 {\it Phys. Rev.} A {\bf 20}  1608.
\bibitem{IJKK} Ingarden R,  Janyszek H,  Kossakowski A and  Kawaguchi 1982 {\it Tensor N.S.} {\bf 37}  105.
\bibitem{Wo} Wootters W 1981 {\it Phys. Rev.} D  {\bf 23}  357.
\bibitem{G1}  Gilmore R 1984 {\it Phys. Rev.} A  {\bf 30 } 1994.
\bibitem{R2}  Ruppeiner G 1985 {\it Phys. Rev.} A  {\bf 32}  3141.
\bibitem{G2} Gilmore R 1985 {\it Phys. Rev.} A  {\bf 32}  3144.
\bibitem{NS}  Nulton J and  Salamon P 1985{\it Phys. Rev.} A  {\bf 31}  2520.
\bibitem{J1}  Janyszek H 1986 {\it Rep. Math. Phys. } {\bf 24}  1; {\it Rep. Math. Phys.} {\bf 24 }11.
\bibitem{JM}  Janiszek H and  Mrugala R 1989{\it Rep. Math. Phys. } {\bf 27}  145.
\bibitem{R}   Ruppeiner G 2010 {\it Am. J. Phys.} {\bf 78}  1170, and references therein.
\bibitem{JM1}Janiszek H and  Mrugala R 1990 {\it J. Phys. A: Math. Theor.} {\bf 23}  467.
\bibitem{BH}  Brody D and  Hook D 2009 {\it J. Phys. A: Math. Theor.} {\bf 42} 023001.
\bibitem{JM2}Janiszek H and  Mrugala R 1989 {\it Phys. Rev.}A  {\bf 39} (1989) 6515.
\bibitem{J2}  Janyszek H 1990 {\it J. Phys. A:Math. } {\bf 23}  477.
\bibitem{BR} D. Brody and N. Rivier 1995 {\it Phys. Rev.}E {\bf 51}  1006.
\bibitem{JJK}  Janke W,  Johnston D and   Kenna R 2004 {\it Physica} A  {\bf 336}  181.
\bibitem{T-B}  Trasarti-Battistoni R  cond-mat/0203536.
\bibitem{PPP}  Portesi M, Plastino A and  Pennini F 2006 {\it Physica }A {\bf 365}  173.
\bibitem{O}    Ohara A 2007 {\it Phys. Lett. } A  {\bf 370}  184.
\bibitem{MH1}  Mirza B and Mohammadzadeh H 2008 {\it Phys. Rev.} E  {\bf 79}  021127;2009 {\it Phys. Rev.} E  {\bf 80}  011132; 2010 {\it Phys. Rev.} E {\bf 82}  031137;
2011 {\it J. Phys. A: Math. Theor.} {\bf 44 } 475003. 
\bibitem{MRU1}  Ubriaco M R 2012 {\it Phys. Lett. } A {\bf 376 } 2899.
\bibitem{MRU2}  Ubriaco 2012 {\it Phys. Lett. } A {\bf 376} 3581.
\bibitem{MRU3}  Ubriaco 2013 {\it Physica} A {\bf 392} 4868.
\bibitem{MRU4}  Ubriaco 2014 {\it Physica} A {\bf 414} 128.
\bibitem{Jimbo} See, for example, Jimbo M ed. 1990 {\em Yang-Baxter equation
in integrable systems}, Advanced series in Mathematical Physics V.10
(World Scientific).
\bibitem{Ch}  Chari V and Pressley A 1994 {\em A Guide to Quantum Groups},
(Cambridge Univ. Press).
\bibitem{CW} See, for example,  Carow-Watamura U,  Schlieker M,  Scholl M and  Watamura S 1990
{\it Z. Phys.} C  {\bf 48} 150 ;  Ogievetsky O,  Schmidke W,  Wess J and  Zumino B 1990
{\it Int. J. Mod. Phys.} A {\bf 6} 3081 .
\bibitem{AV}   Aref'eva I and  Volovich I 1991 {\it Phys. Lett } B  {\bf 264} 62;
  Kempf A 1994 {\it J. Math. Phys.} {\bf  35} 4483 ;
  Brzezinski T and  Majid S 1993 {\it Phys. Lett.} B {\bf 298} 339 ;
  Castellani L 1994 {\it Mod. Phys. Lett. } A {\bf 9} 2835 ;
 Ubriaco M R 1993 {\it Mod. Phys. Lett. } A {\bf 8} 2213 ; 1995 {\bf 10} 2223(E) ;
 Ubriaco M R 1994 {\it Mod. Phys. Lett.} A {\bf 9} 1121 ;
 Sudbery A 1996 {\it Phys. Lett.} B {\bf 375} 75.
\bibitem{I}   Iwao S 1990 {\it Prog. Theor. Phys.} {\bf 83} 363 .
  Bonatsos D,   Argyres E and  Raychev P 1991 {\it J. Phys.} A {\bf 24}
L403;  Capps R 1994 {\it Prog. Theor. Phys.}  {\bf 91} 835.
\bibitem{MRU5}  Ubriaco M R 1996 {\it Phys. Lett. } A {\bf 219 }205 ;
  1998 {\it Phys. Rev.}E  {\bf 57} 179 ; 1998 {\it Phys. Rev.}E {\bf 58} 4191 . 
\bibitem{MRU6}  Ubriaco M R 1997  {\it Phys. Rev.}E {\bf 55} 291 .
\bibitem{MN}  Meusburger C and  Noui K 2010 {\it Adv. Theor. Math. Phys.}
{\bf 14 } 6  1651,  and references therein.
\bibitem{KWL} Korbicz J,  Wehr J and   Lewenstein M 2009 {\it J. Math. Phys.} {\bf 50} 062104.
\bibitem{VWZ}  Vokos S,   Zumino B and  Wess J 1989 {\em Symmetry in Nature} (Scuola Normale Superiore Publ.,
Pisa, Italy).
\newpage
\bibitem{AAA} Algin A,  Arik M and  Arikan A  2002 {\it Phys. Rev.}E {\bf 65} 026140 .
\bibitem{F} Feder J 1988 {\em Fractals} (Plenum, New York).
\bibitem{BD}  B\"{u}y\"{u}kkili\c{c} F and  Demirhan D 1993 {\it Phys. Lett. } A {\bf 181}  24.
\bibitem{BDA}   B\"{u}y\"{u}kkili\c{c} F,  Demirhan D and  G\"{u}le\c{c} A 1995 {\it Phys. Lett. } A {\bf 197}  209.
\bibitem{T}  Tsallis C 1988 {\it J. Stat. Phys.} {\bf 52}  479.
\bibitem{MRU7}  Ubriaco M R 1999 {\it Phys. Rev.}E {\bf 60 } 165.

\end{thebibliography}
\end{document}